\documentclass[manuscript]{aastex}

\usepackage{rotating}
\usepackage{tikz}
\usepackage{amsmath}
\usepackage{gensymb}
\usepackage{booktabs}
\usepackage{pdflscape}
\usepackage{color,soul}
\usepackage{hyperref}

\slugcomment{The Astronomical Journal}

\shorttitle{SW3 73P-C}
\shortauthors{Graykowski et al.}

\begin{document}

\title{Fragmented Comet 73P/Schwassmann-Wachmann 3}


\author{Ariel Graykowski$^{1}$ and  David Jewitt$^{1,2}$
}

\affil{$^1$Department of Earth, Planetary and Space Sciences,
UCLA, 595 Charles Young Drive East, Los Angeles, CA 90095-1567\\
$^2$Department of Physics and Astronomy, University of California at Los Angeles, 430 Portola Plaza, Box 951547, Los Angeles, CA 90095-1547\\
}

\email{ajgraykowski@ucla.edu}

\begin{abstract}

Comet 73P/Schwassmannâ-Wachmann 3  has been observed to fragment on several occasions, yet the cause of its fragmentation remains poorly understood. We use previously unpublished archival Hubble Space Telescope data taken in 2006 to study the properties of the primary fragment, 73P-C, in order to constrain the potential fragmentation mechanisms. Currently the literature presents a wide range of measured rotational periods, some of which suggest that the nucleus might have split due to rotational instability. However, we find the most likely value of the rotation period to be 10.38 $\pm$ 0.04 hours (20.76 $\pm$ 0.08 hours if double-peaked), much longer than the critical period for rotational instability for any reasonable nucleus density and shape, even in the absence of tensile strength. We also find strong, cyclic photometric variations of about 0.31 $\pm$ 0.01 magnitudes in the central light from this object, while similar variations with a smaller range are apparent in the surrounding dust coma.  These observations are compatible with rotational modulation of the mass loss rate and with dust having a mean outflow speed of $107 \pm 9$ m s$^{-1}$. Finally, we also estimate the radius of the nucleus to be 0.4 $\pm$ 0.1 km accounting for dust contamination and assuming a  geometric albedo of 0.04.

\end{abstract}


\section{Introduction}
Comet 73P/Schwassmann-Wachmann 3 (hereafter 73P) is a Jupiter family comet with a semimajor axis $a$ = 3.063 AU, orbital  inclination $i$ = 11.4$\degree$, and eccentricity $e$ = 0.692. The comet was first reported to have fragmented into at least four pieces on September 12, 1995 (Crovisier et al., 1995), shortly before its perihelion passage at 0.94 AU on September 22. While fragments A and D were not seen during its following apparition, fragments B and C are long-lived, continuing to appear in subsequent apparitions every 5.36 years. Fragment 73P-C is the brightest of the fragments and is also the leading fragment in the orbit. These characteristics generally describe the primary fragment of a split comet according to Boehnhardt (2004), so 73P-C is therefore considered the primary nucleus. While its 2001 apparition provided a less than ideal observing geometry, the comet approached to within 0.08 AU of the Earth in 2006, providing an excellent opportunity for high resolution observations with the Hubble Space Telescope (HST). 73P fragmented again during this apparition, releasing fragments G, H, J, L, M and N, with fragments B and G famously shedding dozens of smaller pieces (Fuse et al., 2007 and Weaver et al., 2008). 

Potential causes of cometary fragmentation include tidal disruption (Asphaug and Benz 1994), rotational instability (Jewitt 1997), internal build-up of thermal gas pressure (Samarasinha 2001), and impact induced fragmentation (Sekanina 1997 and Boehnhardt 2004). In the case of 73P-C, tidal disruption can be discounted since the orbit of the comet does not pass within the Roche spheres of any of the major planets or the Sun.  Impacts are intrinsically unlikely and offer an even less credible explanation given that the comet has exhibited multiple break-up episodes in different orbits.  Estimates of the rotational period of 73P-C (c.f.~Table \ref{past}) are widely spread over the range from 3.019 hours (Drahus et al., 2010) to 27.2 hours (Storm et al.~2006).  The lower end of this range is suggestive of fragmentation due to rotational instability  (Marzari et al.~2011). However, the upper end of the range is completely inconsistent with this possibility. 

In this paper, our motivation is to characterize 73P using the highest resolution data and to obtain a deeper understanding of how 73P fragmented. We focus on 73P-C and examine HST images from April, 2006 and measure the rotation period of 73P-C in order to reduce the ambiguity behind the cause of its fragmentation. 

\section{Observations}

The images were taken under HST program GO 10625, with P. Lamy as the principal investigator using the Advanced Camera for Surveys (ACS) in the High Resolution Channel (HRC). The field of view of the HRC is $29\arcsec \times 26\arcsec$ with a pixel scale of $0.028\arcsec \times 0.025\arcsec$/pixel (Ford et al., 1998 and Ryon et al. 2019).  The observations started on 2006 April 10, 21:35 UT and ended 2006 April 11, 19:06 UT, providing a span of $\sim$21.5 hours. A total of eight orbits were scheduled to observe 73P-C. Six of the eight orbits consisted of observations that were identical in filters, the number of images taken in each filter, and the exposure times. In each orbit, two images were taken with the F475 filter, two with F555, six with F606, and four with F625. The first orbit is almost identical, but features shorter exposure times for the two images taken in the F475 filter. The eighth orbit also features shorter exposure times of the two images taken in F475, as well as the two images in F555 and the four images in F625 in order to provide time for two additional images taken in the F814 filter. We focused our analysis on the filters consistent between each orbit. Orbits four and seven are not available on the Hubble Legacy Archive, leaving a total of 84 DRZ images (calibrated, geometrically-corrected, dither-combined images created by AstroDrizzle in units of electrons/s) that were analyzed in this work (Table \ref{filters}). The observing geometry is listed in Table (\ref{geometry}) for the two days on which the comet was observed.

\subsection{Nucleus Measurements}
\begin{figure}
\epsscale{1}
\plotone{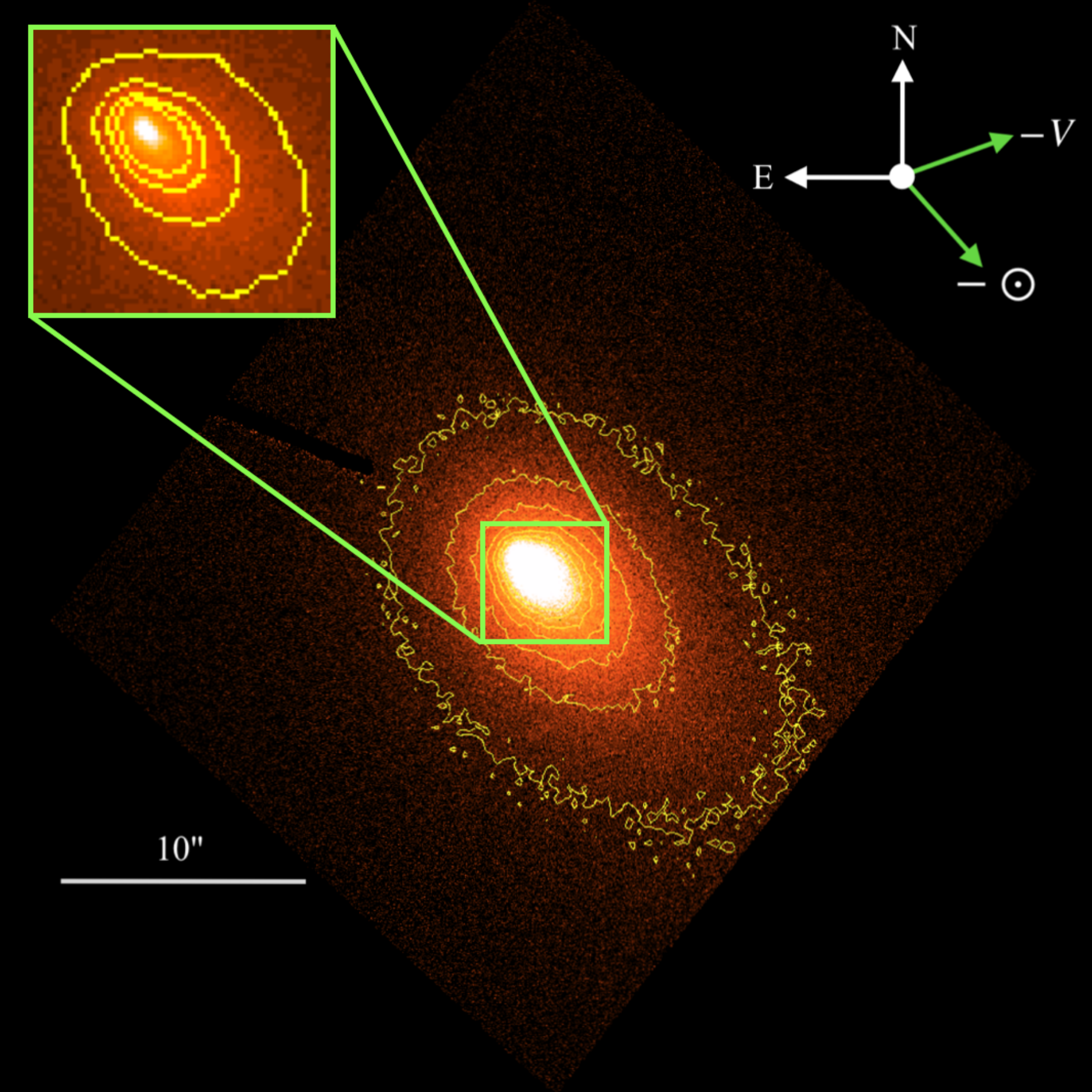}

  \caption{Sample single image of 73P taken with HST's ACS HRC instrument in the F555 filter with a 20 second exposure on April 10, 2006. The pixel scale is 0.028\arcsec $\times$ 0.025\arcsec~/ pixel. At a distance of 0.288 AU from the Earth, 1$\arcsec$ corresponds to 208.9 km and one pixel subtends $\sim$5.2 km. The 5.2\arcsec $\times$ 4.0\arcsec~inset magnifies the area around the nucleus stretched to minimize saturation. The projected antisolar direction and negative velocity  are represented by the vectors labeled -$\odot$ and $-V$ respectively. \label{piccy}}
\label{contour}
\end{figure}

The morphology of  the parent body, 73P-C, can be seen in Figure (\ref{contour}). It displays an obvious coma elongated towards the southwest direction, approximately aligned with the antisolar direction (marked -$\odot$). We used aperture photometry to measure the apparent magnitude in the 84 images and accounted for the zeropoint of each filter given by \url{http://www.stsci.edu/hst/acs/analysis/zeropoints/old_page/localZeropoints}. For the nucleus measurements, we chose an aperture radius of $\sim$0.13$\arcsec$ ($\sim$FWHM of the PSF), which corresponds to a distance of 27 km at the comet. We then subtracted the background flux caused by the sky and coma through an annulus contiguous with the aperture and extending out another $\sim$0.26$\arcsec$. In order to analyze the data from each filter all together, we calculated the average colors with respect to filter F555 (equivalent to the Johnson V filter) and used the colors to shift the magnitudes in each filter such that they are equivalent to the F555 filter (Table \ref{colors}). The nucleus aperture magnitudes are plotted as a function of time in Figure (\ref{nuclight34}).

\subsection{Coma Measurements}
\begin{figure}
\epsscale{1}
\plotone{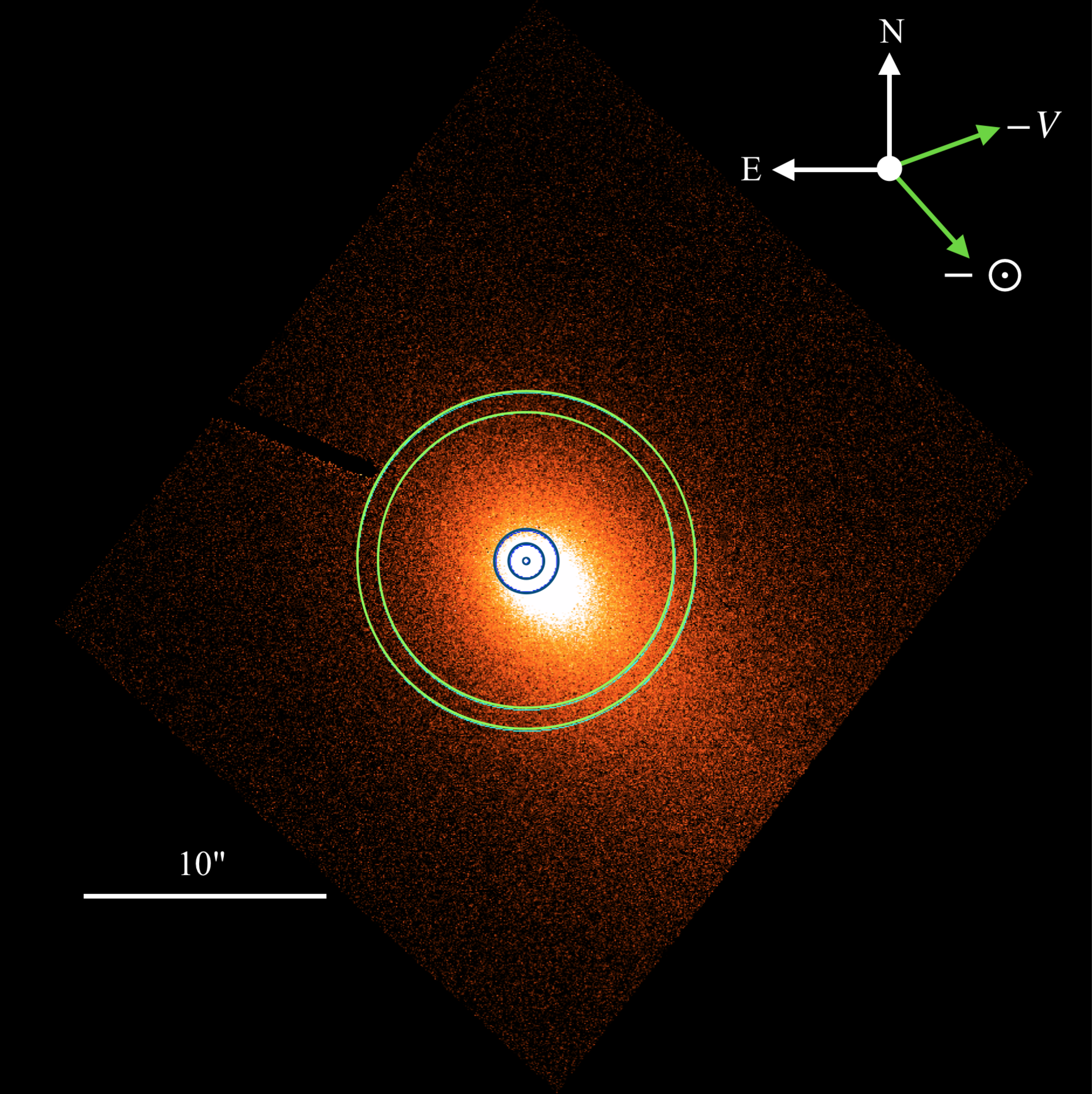}
  \caption{Same as Figure (\ref{piccy}) but with blue circles  to show photometric apertures with radii 27, 136, and 245 km and green circles to represent the inner and outer edges of the background annulus corresponding to radii $\sim$1190 and 1360 km. }
\label{aperture}
\end{figure}

\begin{figure}
\epsscale{0.85}
\plotone{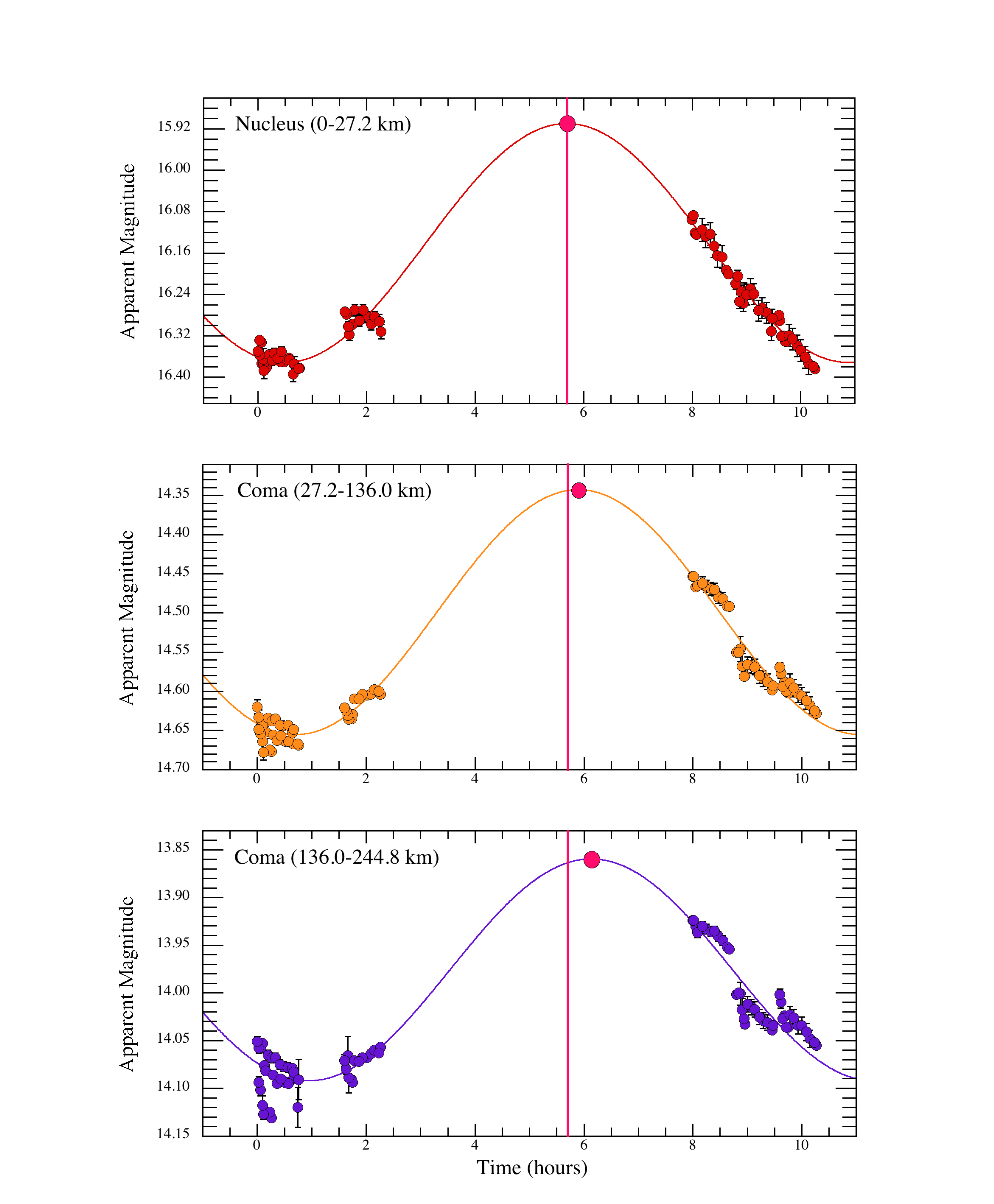}
  \caption{Phase folded lightcurves of the nucleus (top), the coma with inner and outer annulus radius of 27 and 136 km (middle), and the coma with an inner and outer annulus radius of 136 and 245 km (bottom). The magenta, vertical line is placed at the peak of the nucleus lightcurve, and is unmoved in the subsequent coma lightcurves. The magenta circle marks the fitted peak of each lightcurve. A phase shift is clearly visible from the nucleus to greater distances in the coma.}
\label{phasefoldshift}
\end{figure}

Next, we compared the lightcurve of the central aperture (containing the nucleus and near-nucleus dust) to lightcurves of the coma annuli (containing only dust). We used the same aperture radius to obtain the lightcurve for the nucleus. We then measured the flux within the coma within two annuli extending from $\sim$0.13$\arcsec$ to $\sim$0.65$\arcsec$ and from $\sim$0.65$\arcsec$ to $\sim$1.17$\arcsec$. Instead of subtracting the background flux from an annulus immediately outside these regions, we only subtracted the flux from the distant sky background. For this purpose, we experimented to select an annulus with inner and outer radii of $\sim$5.7$\arcsec$ and $\sim$6.5$\arcsec$ respectively, and subtracted this flux from all measurements. The aperture and annuli are drawn on an image of the comet in Figure (\ref{aperture}). Colors were again calculated in the coma regions with respect to filter F555 as listed in Table (\ref{colors}), with error estimates representative of the signal-to-noise ratio (SNR). The SNR decreased further out in the coma, where the sky background had a larger effect.The resulting apparent magnitudes are plotted in Figure (\ref{phasefoldshift}). 

\section{Results}

\subsection{Rotational Period of the Nucleus}

\begin{figure}
\epsscale{1}
\plotone{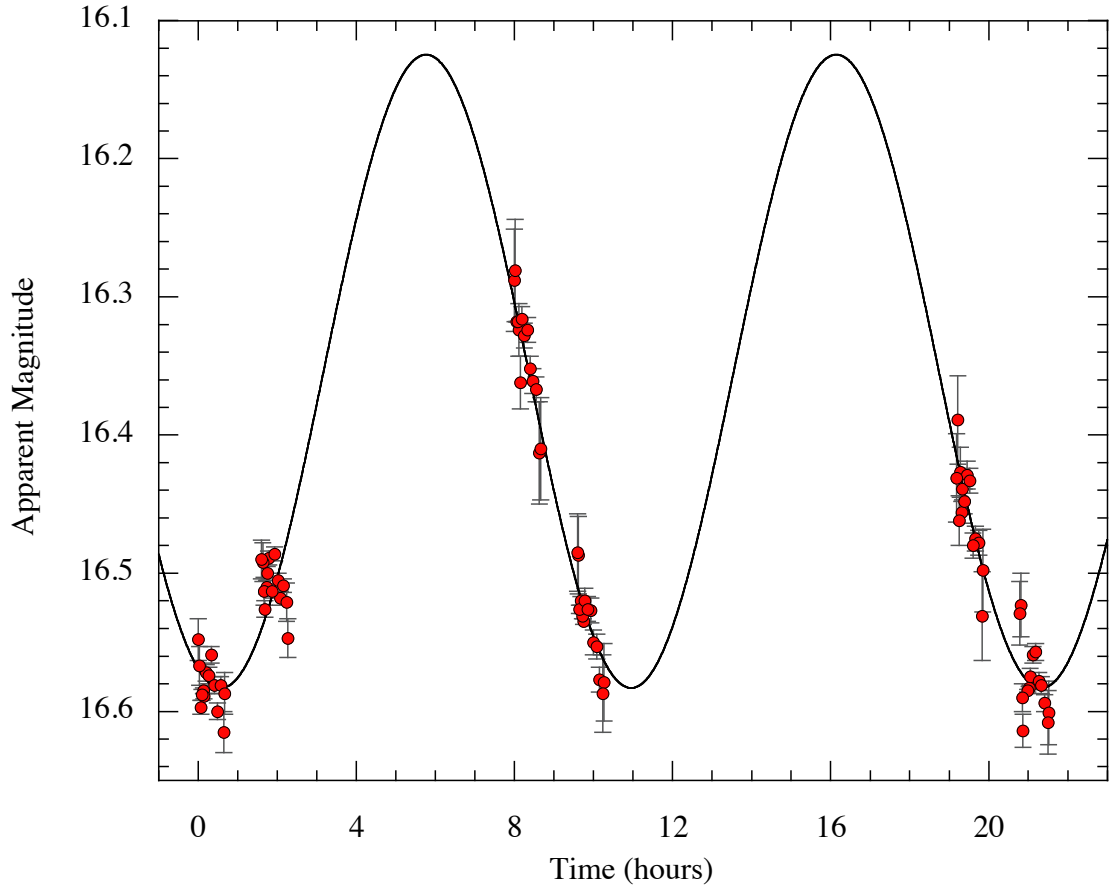}
  \caption{73P-C apparent magnitude versus time plotted as red circles with 1$\sigma$ error bars. The solid line shows a best-fit sinusoidal lightcurve with a period $P = 10.38 \pm 0.04$ hours and a rotation period $2P = 20.76 \pm 0.08$ hours.}
\label{nuclight34}
\end{figure}

	We measured the rotation period of the nucleus by fitting a lightcurve to the time-varying magnitudes. For convenience of presentation, the time plotted in Figures (\ref{nuclight34}) and (\ref{phasefoldshift}) was calculated from 
	
\begin{equation}
T = (JD - 2453836.399) \times 24
\label{time}
\end{equation}

\noindent where $T$ is time in hours and JD is the Julian date obtained from the image header. We used IRAF's phase dispersion minimization (PDM) program to search for likely periods in the data, and used these values as a guide to create the best fit sinusoid curves by least squares. We fit the data from each filter separately, as well as combined by converting the F475, F606, and F625 filters to the F555 filter according to the colors in Table (\ref{colors}). We found that all five of these fits resulted in the same period within two standard deviations. Therefore, for simplicity, we only show the fit for the combined data in this work. Figure (\ref{nuclight34}) shows the best fit for the lightcurve of the nucleus. This sinusoidal  least-squares fitted curve  has a single-peaked period $P$ = 10.38 $\pm$ 0.04 hours. The lightcurves of most solar system small bodies are dominated by variation in the cross-section due to aspherical shape, rather than by surface albedo variations (Burns and Tedesco, 1979). In these cases the  lightcurves are doubly periodic (two maxima and minima per rotation) owing to rotational symmetry. For 73P-C, this would imply a rotational period $2P$ = 20.76 $\pm$ 0.08 hours.  This relatively long period is consistent with a lower limit $P >$ 10 hours set using radar observations by Nolan et al.~(2006), but disagrees with shorter periods obtained using less direct methods (Table \ref{past}).  Incomplete sampling of the lightcurve leads to aliasing in the period determination and, therefore, we cannot exclude the possibility that other periods might fit the data.  The shortest plausible period from our PDM analysis is $P$ = 5.01 hours and thus we can rule out shorter periods proposed by Drahus et al. (2010), Toth et al. (2006 and 2008) listed in Table (\ref{past}).

%Incomplete sampling of the lightcurve leads to aliasing in the period determination and, therefore, we cannot exclude the possibility that other periods might fit the data. An absolute lower limit is set by the monotonic  portion of the lightcurve between 8.00 and 10.27 hours (spanning 2.27 hours). For a rotating body, the lightcurve must return back to the brightest magnitude after this descent. Therefore, at the very least, the period of the lightcurve must be at least twice that of the observed descent, giving a minimum period of 4.53 hours. This is important as it eliminates the possibility of rotational instability as the break up mechanism, as will be discussed further in the Discussion section.

\subsection{Dust Outflow Speed}

\begin{figure}
\epsscale{1}
\plotone{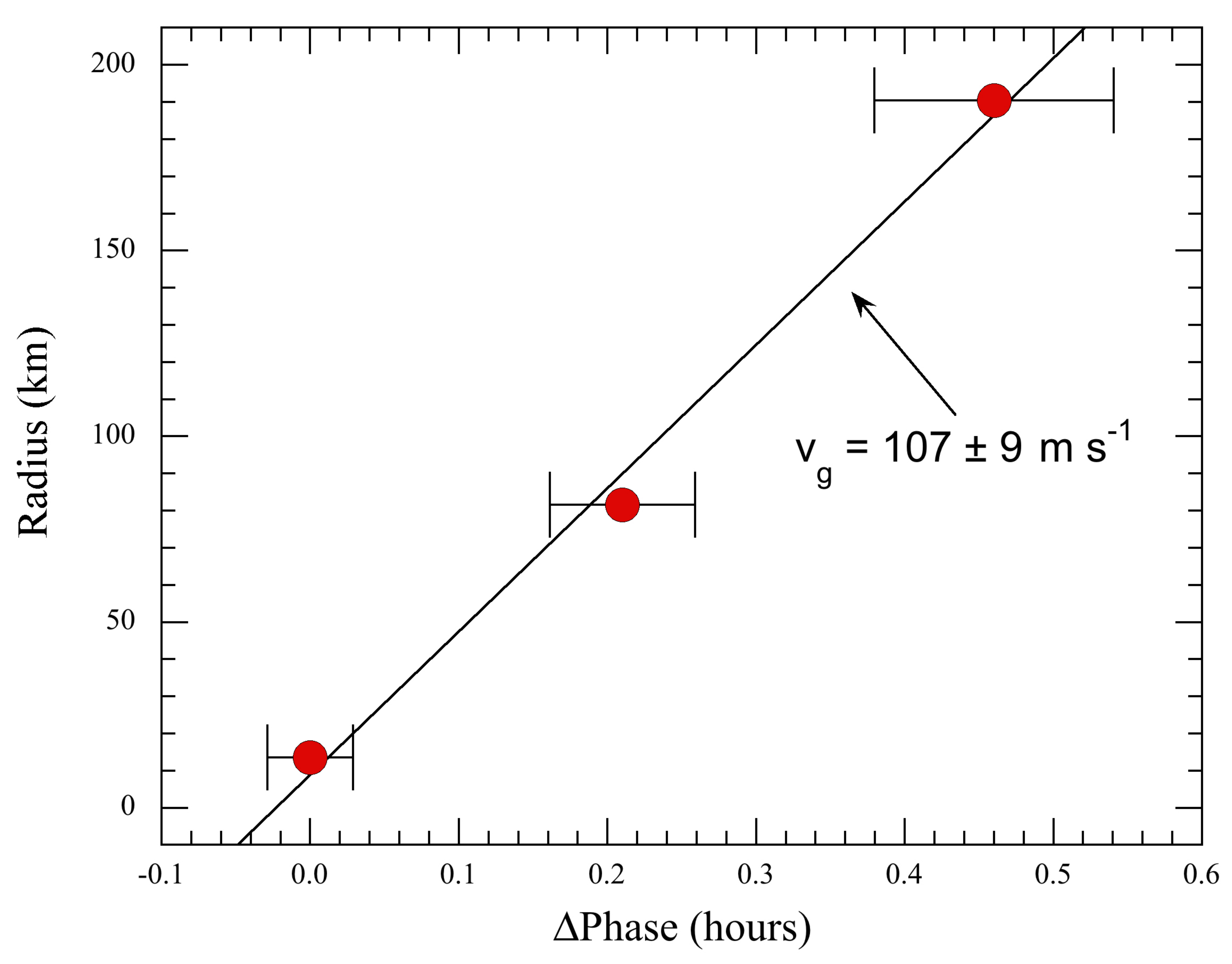}
  \caption{The relative phase shift plotted against the distance from the center of the nucleus, where the reference phase measured from the lightcurve of the nucleus is set to zero. From left to right, the red points then represent the relative phase at the nucleus, 82 km from the nucleus into the coma, and 190 km from the nucleus into the coma. The error bars represent the one sigma error on the phase obtained from each fit. From this, we obtain a dust speed $v_g =$ 107 $ \pm$ 9 m $s^{-1}$.}
\label{phaseshiftspeed}
\end{figure}

\begin{figure}
\epsscale{1}
\plotone{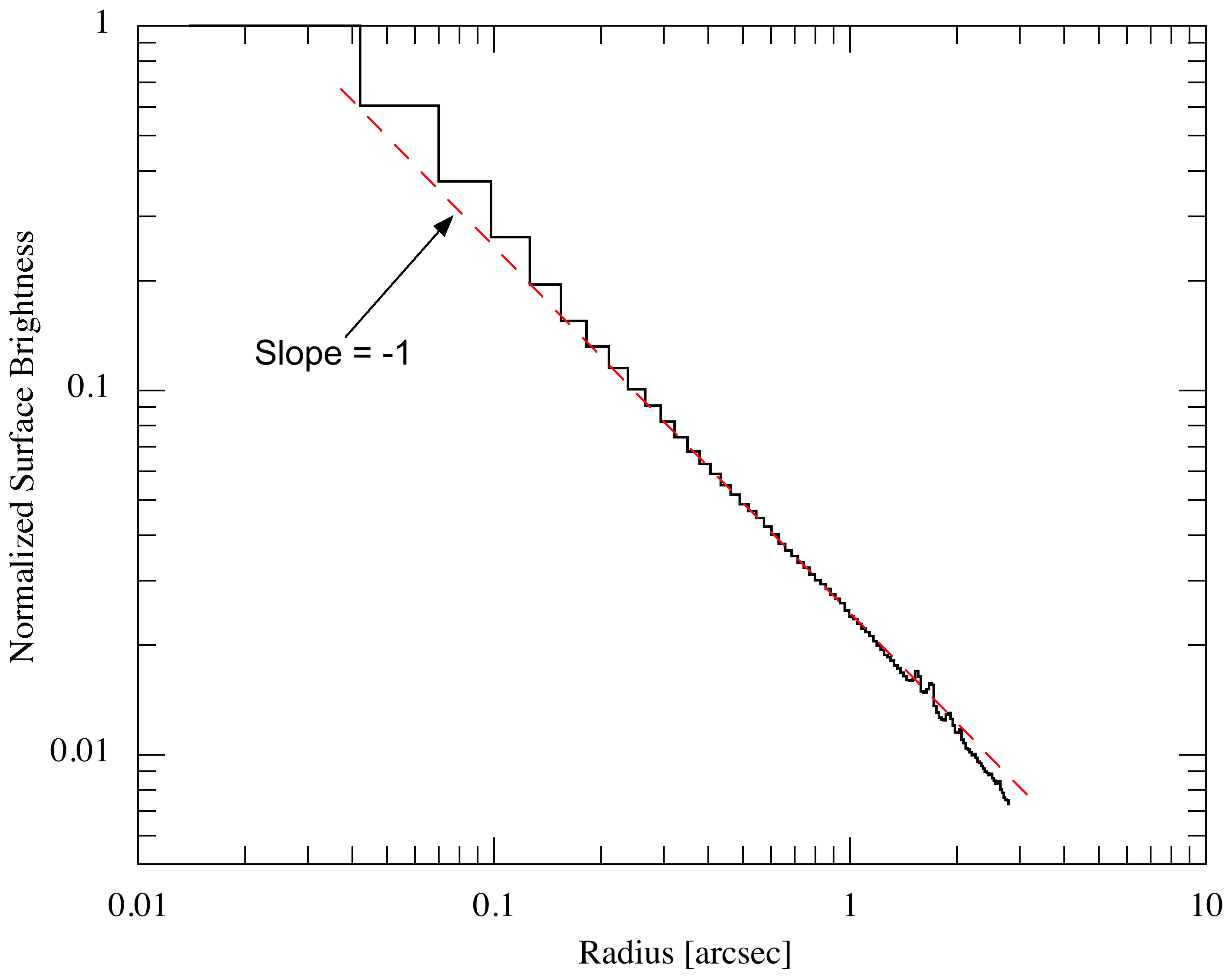}
  \caption{The surface brightness profile of 73P. The red line represents a surface brightness profile $\propto r^{-1}$, where $r$ is the radial distance from the center of the nucleus. The surface brightness profile follows this line closely except $r \lesssim$ 0.15\arcsec~where the PSF has an effect, and $r \gtrsim$ 1.5\arcsec, where the background subtraction systematics are important.}
\label{slope}
\end{figure}

We next repeated the analysis for the lightcurves of the nucleus and coma regions measured from the sky-subtraction technique. Specifically, we aim to detect a phase lag in the lightcurves from the three apertures caused by the finite speed of outflow of the dust. Keeping the same period found in our best fit of the nucleus lightcurve in the previous section, we fitted curves with the lightcurve period $P$ =  10.38$\pm$0.04 hours to the phase-folded data in Figure (\ref{phasefoldshift}). The figure shows that there is a phase lag  from the nucleus to the dust regions of the coma. The fitted phase lags are listed in Table (\ref{params34}). From the central aperture to the inner coma annulus there is a $\Delta T$ = 0.21 $\pm$ 0.05 hour lag while from the center to the outer coma annulus the lag is $\Delta T$ = 0.46 $\pm$ 0.08 hour.  Figure (\ref{phaseshiftspeed}) shows the  phase lag versus the effective radius of each photometry annulus, computed by determining the center of light of the photons hitting the CCD in each aperture or annulus according to the surface brightness profile plotted in Figure (\ref{slope}). With a surface brightness  $\propto \theta^{-1}$ (where $\theta$ is the radial distance from the center), the effective radius, $\theta_e$, is given simply  by $\theta_e = (\theta_i + \theta_o)/2$, where  $\theta_i$ and $\theta_o$ are the inner and outer radii of each annulus. The speed of the dust, $v_g$, is given by the gradient of a straight line fitted to the data in Figure (\ref{phaseshiftspeed}).   We find $v_g =$ 107 $\pm$ 9 m/s.  At this speed, dust released from the nucleus would escape from the central photometry aperture of projected radius $\theta_o$ = 27 km (Table \ref{params34}) in a time $t \sim \theta_o/V_g \sim$ 250 s.  This time is very short compared to the nucleus rotational period and, therefore, we can neglect the effects of averaging in the interpretation of the lightcurve data. %Thus, we can immediately discount the possibility that the lightcurve variations in 73P-C are modulated by cyclic variations in the amount of dust in the coma and instead reflect rotation of the nucleus, as we earlier assumed.

\subsection{Size of the Nucleus}

We converted from apparent to absolute magnitude, $H_V$, defined as the magnitude corrected to unit heliocentric and geocentric distance ($r_H$ and $\Delta$, respectively) and to phase angle  $\alpha$ = 0\degr.  For the apparent magnitude, $m_V$, the correction is 

\begin{equation}
H_V= m_V - 5\log_{10}(r_H \Delta) - f(\alpha)
\label{abs}
\end{equation}

\noindent where $f(\alpha)$ is the phase function representing the angular dependence of the scattered sunlight at phase angle $\alpha$ in degrees. We assume a linear phase function 

\begin{equation}
f(\alpha) = (0.046 \pm 0.017) \alpha
\label{phasefunc}
\end{equation}

\noindent based on measurements of 37 Jupiter family comets reported by Kokotanekova (2017). We find the absolute magnitude of 73P-C to be $H_V$ = 17.3 $\pm$ 0.1.  

We estimated the cross-section of the nucleus using

\begin{equation}
C_e=\dfrac{{\pi}(2.25\times10^{16})}{p_V}10^{-0.4[H_V-V_{\odot}]}
\label{crosssection}
\end{equation}

\noindent where $p_V$ is the geometric albedo, which we assume to be 0.04 (Hartmann et al.~1987, Lamy et al.~2004, Fern{\'a}ndez et al. 2013), $H_V$ is the absolute magnitude and $V_{\odot}\sim$ -26.77 is the apparent magnitude of the Sun. The effective nucleus radius is then $r_n=\sqrt{(C_e / \pi)}$ and we find $r_n = 1.2\pm0.1$ km. This is consistent with the pre-breakup radius $r_n$  $\sim$ 1 km as reported by Sekanina~(1989) and $r_n$ $\sim$ 1.1 km as reported by Boehnhardt et al.~(1999). In both of these studies, as well as our calculation of the radius, coma contamination was not accounted for, and therefore these values all represent upper limits to the size of the nucleus. 

Toth et al.~(2005) accounted for coma contamination by modeling the expected brightness profile of the coma and nucleus, and calculating the residual flux associated with just the nucleus. From this, Toth et al.~(2005) find that, as of 2001 November 26, 73P-C had decreased in radius to $r_n$ = 0.68 $\pm$ 0.04 km (i.e.~to only 25$\%$ of the original volume). However, the smaller size is likely a result of accounting for dust contamination rather than the nucleus physically shrinking over time. We conducted a similar analysis by extrapolating the surface brightness profile in the coma to the center aperture. We subtracted the estimated flux of the coma within the center aperture from the flux value we measured using aperture photometry. The residual flux represents that of the isolated nucleus. In the measured region of the coma, Figure (\ref{slope}) shows the surface brightness profile $\propto r^{-1}$, where $r$ is the radial distance from the center of the nucleus. This implies that the integrated flux within an annulus is simply proportional to the width of the annulus, i.e.~flux $\propto r_o-r_i$. Therefore, if there is no nucleus present, we expect to find flux ratios in the different apertures $A1/A2$ = 27/(136-27) = 0.25 and $A1/A3$ = 27/(245 - 136) = 0.25 (c.f.~Table \ref{params34}).  Instead, the average flux measured through the center aperture is $\sim29\%\pm2.5\%$~ of the average flux in the indicated coma annuli. The residual flux corresponds to an apparent magnitude $m_V\sim18.7\pm0.7$, absolute magnitude $H_V\sim19.7 \pm 0.7$, and radius $r_n\sim0.4\pm0.1$ km. This estimate (diameter 0.8$\pm$0.2 km)  is comparable, within the uncertainties, to Arecibo (12.6 cm) and Goldstone (3.5 cm) radar data reported to show that fragment C is at least 1 km in diameter (Nolan et al. 2006).

\section{Discussion}

\subsection{Stability of the Nucleus Rotation}

Is rotational instability a plausible mechanism of fragmentation for 73P? We compare the minimum period we derived with the critical rotational period of the body given by

\begin{equation}
P_C=k\left[ \frac{3\pi}{G\rho} \right]^{\frac{1}{2}}
\label{rotinst}
\end{equation}

\noindent where $G$ is the gravitational constant, $\rho$ is the density, $k$ is a dimensionless constant that depends on the shape of the body. For a sphere, $k$ reduces to 1. We assume that a prolate shaped nucleus generates the lightcurve, so $k=a/b$ where $a$ $>$ $b$. We also assume a density $\rho$ = 600 kg/m$^3$, as representative for comets (Britt et al.~2006). With photometric variations of $\sim0.31 \pm 0.01$ magnitudes in the nucleus, we infer $k=1.33 \pm 0.01$, and a critical period $P_C = 5.67 \pm 0.04$ hours for a prolate spheroid of this density, which is far shorter than our 10.38 hour best-fit single-peaked lightcurve period.  Therefore, we can confidently rule out the possibility of break up of a strengthless nucleus due to rotational instability.

\subsection{Coma Dust Speed and Particle Size}

From the measured speed of the dust in the coma, we can estimate the average dust grain size and mass loss rate from the nucleus. We assume an inverse relationship between ejection velocity and grain size 

\begin{equation}
v = v_0 \beta^{\frac{1}{2}}
 \label{velocity}
\end{equation}

\noindent where $\beta$ is the ratio of the acceleration due to solar radiation pressure to the acceleration due to solar gravity and $v_0$ is the reference velocity for a particle 1 $\micron$ in radius (Lisse et al.~1998; Reach et al.~2000; Ishiguro et al.~2007) and $\beta \sim 1/a$ where $a$ is the grain radius in $\micron$. We then can evaluate the average velocity of particles by weighting the velocity by the size distribution and the cross-section of the particles. The average velocity is

\begin{equation}
\bar{v} = \frac{\int_{a_{min}}^{a_{max}} v_0 \pi a^{\frac{3}{2}} n(a) da}{\int_{a_{min}}^{a_{max}}  \pi a^2 n(a) da}
 \label{averagevelocity}
\end{equation}

\noindent where n(a)da is the size distribution of dust. This power law distribution is defined as 

\begin{equation}
n(a) da = \Gamma a^{-\gamma} da
 \label{averagevelocity2}
\end{equation}

\noindent where $\Gamma$ is a constant that does not affect the resulting velocity, and we take $\gamma = 3.5$, as found in other active bodies (e.g.~Jewitt et al.~2014). Then, assuming $a_{max}$ $\gg $ $a_{min}$, the resulting average velocity is 

\begin{equation}
\bar{v} = \frac{v_0 }{2 \sqrt{a_{min}}}
 \label{resultvelocity}
\end{equation}

\noindent with $a_{min}$ expressed in \micron. 

Equation (\ref{resultvelocity}) shows that, with these assumptions, the average velocity only depends on the minimum particle size. The smallest particles that can be seen in the visible spectrum are limited by diffraction. The scattering efficiency of a particle depends on the size parameter, $x = 2\pi a/\lambda$, where $\lambda$ is the wavelength of observation. The scattering efficiency approaches 0 as $x \rightarrow$ 0, and oscillates around and approaches unity when  $x~\textgreater$ 1 (Van de Hulst 1957). In the visible spectrum ($\lambda \sim$0.5 \micron), $x =$ 1 corresponds to $a \sim 0.1$ \micron. These tiny particles are  likely to be dynamically well coupled to the gas and ejected with a velocity that is comparable to the average thermal velocity, given by integrating the Maxwell Boltzman distribution of particle speeds; 

\begin{equation}
v_{th} = \sqrt{\frac{8 k_B T}{\pi m}}.
 \label{gasvelocity}
\end{equation}

\noindent Here the spherical blackbody temperature at $\sim$1.23 AU from the Sun is $T\sim251$ K, $k_B$ is the Stefan-Boltzmann constant, and $m$ is the mass of an H$_2$O molecule. Equation (\ref{gasvelocity}) gives $v_{th}$ $\sim$ 550 m s$^{-1}$, which is our best estimate of the outflow speed in gas. If we assume that the smallest particles, $a \sim$ 0.1 \micron\, are dynamically well coupled to the gas and traveling at $v_{th}$, then the mean speed $\bar{v}\sim$107 $\pm$ 9 m s$^{-1}$ corresponds to effective particle radius $a \sim$ 3 \micron,  by Equation (\ref{resultvelocity}).

This particle size estimate is clearly approximate, given that we do not know the actual gas flux from the nucleus and that $v_{th}$ is only an approximation to the outflow speed.  However, the sunward extent of the coma provides an independent estimate of the particle ejection speed (Jewitt 1991).
A particle launched towards the Sun and experiencing a constant anti-solar acceleration of magnitude $\beta g_{\odot}$ will reach a turn-around distance given by 

\begin{equation}
L = \frac{v^2}{2 \beta g_{\odot}}.
\label{L}
\end{equation}

\noindent Substituting $v = v_0 \beta^{1/2}$, Equation (\ref{L}) simplifies to $L = v_0^2/(2 g_{\odot})$, where $g_{\odot} = 4\times10^{-3}$ m s$^{-2}$ is the solar gravity at $r_H$ = 1.23 AU.  We see from Figure (\ref{contour}) that the sunward angular extent of the coma is $\sim$10\arcsec~corresponding to 2$\times10^6$ m at the comet, neglecting the effects of projection.  We infer $v_0 \sim (2 g_{\odot} L)^{1/2} \sim$ 125 m s$^{-1}$, corresponding to the speed of a 1 \micron\ particle.  This is close enough, given the many approximations involved, that we consider this independent estimate to be in strong support of the result from aperture photometry.

\subsection{Mass Loss Rate}

 With the effective velocity in-hand, a simple mass loss rate estimate can be obtained from

\begin{equation}
\frac{dM}{dt}=\frac{4 \rho \bar{a} C_e \bar{v}}{3d}
 \label{massloss}
\end{equation}

\noindent where $\rho$ is the dust grain density, which we assume again to be 600 kg m$^{-3}$, $\bar{a}$ is the average particle radius, $C_e$ is the total cross section of all the particles in the measured portion of the coma as given by Equation (\ref{crosssection}), $\bar{v}$ is the average velocity of particles in the coma, and $d$ is the distance traveled. For $d$, we use the distance from the nucleus to the midpoint of the outer defined region of the coma, $d \sim$190 km. Substituting into Equation (\ref{massloss}) gives the order of magnitude average mass loss rate $dM/dt$ $\sim$ 50 $\pm$ 17 kg s$^{-1}$ where the error is largely dominated by the error on particle size. Finally, we solved the energy balance equation for a patch of perfectly absorbing ice exposed at the subsolar point on the nucleus of 73P. The equilibrium mass loss rate at $r_H$ = 1.23 AU  is $f_s=2.9\times10^{-4}$ kg m$^{-2}$ s$^{-1}$ (ice temperature 202 K). The measured sublimation rate can thus be supplied by a circular patch of area 

\begin{equation}
\pi r^2=\frac{1}{\delta f_s}\frac{dM}{dt}
 \label{activepatch}
\end{equation}

\noindent where $\delta$ is the dust-to-ice mass ratio. If $\delta$ is equal to unity, then $\pi r^2 \sim 0.17 \pm 0.05$ km$^2$ and $r\sim 0.2$ km. On a 0.4 km radius spherical nucleus, this patch corresponds to an active fraction $f_A=\pi r^2 / 4 \pi r_n^2 \sim 0.08$. Published estimates of $\delta$ vary considerably, but there is now a consensus that $\delta > 1$. For example, Reach et. al (2000) found $\delta \sim$ 10-30 in 2P/Encke, while Fulle et al. (2017) find $\delta = 7.5$ for Comet 67P/Churyumov-Gerasimenko. These larger values would imply an active patch of radius $r\sim$0.04-0.08 km on 73P-C. Evidently, very localized activity on 73P can drive the coma.

\section{Conclusion}

We analyzed 84 images of comet 73P/Schwassmann-Wachmann 3 taken with HST on UT 2006 April 10 and 11 to find

\begin{itemize}
\item The best-fit lightcurve period is $P$ = 10.38 $\pm$ 0.04 hours (2$P$ = 20.76 $\pm$ 0.08 hours for a double-peaked lightcurve, as expected for an irregularly shaped body). This eliminates the possibility of rotational instability as a fragmentation mechanism, because the rotation period is above the critical period for breakup for any reasonable density.

\item Accounting for dust contamination, our best estimate of the absolute magnitude is $H_V\sim19.7\pm0.7$ corresponding to a nucleus radius $r_n\sim0.4\pm0.1$ km, assuming an albedo of 0.04.

\item  Phase-lagged brightness variations in the coma show that the dust outflow speed is $v_g = 107 \pm 9$ m s$^{-1}$, corresponding to average dust particle radius  $\bar{a}$ $\sim$ 3 $\pm$ 1 \micron.  The mass loss rate, $dM/dt$ $\sim$ 50 kg s$^{-1}$, is consistent with sublimation of an exposed,  ice patch of radius only $\sim$0.2 km corresponding to $\sim 8\%$ of the nucleus surface.

\end{itemize}

\acknowledgments

We thank Jessica Agarwal, Jing Li, and Max Mutchler for reading the manuscript. This work was supported by the HST's Archival Research Program, proposal \#15301 awarded to D.J. Also, A.G. acknowledges her NESSF grant \#18-0217 for providing funding to accomplish this work. This work was based on observations made with the NASA/ESA Hubble Space Telescope, and obtained from the Hubble Legacy Archive, which is a collaboration between the Space Telescope Science Institute (STScI/NASA), the Space Telescope European Coordinating Facility (ST-ECF/ESA) and the Canadian Astronomy Data Centre (CADC/NRC/CSA).

\clearpage

\begin{deluxetable}{lll}
\tabletypesize{\scriptsize}
%\rotate
\tablecaption{ Reported Periods of 73P-C
\label{past}}
\tablewidth{0pt}
\tablehead{  Rotation Period (Hours) & Author(s) & Method  }

\startdata

 3.019 $\pm$ 0.001,   & Drahus et al.~(2010)& HCN production rate. The method produces many possible \\
   3.349 $\pm$ 0.002, && periods, including solutions in the range of 10.174 to 13.567 \\
    3.392 $\pm$ 0.002 &&  hours, but is insensitive to periods $>$14 hours.\\
    
3.2 $\pm$ 0.2 &Toth et al.~(2006)& Photometry\\

3.5 - 4.0 &Toth et al.~(2008)& Photometry\\

8.8 $\pm$ 0.3,& Storm et al.~(2006)& Dust morphology\\
13.2 $\pm$ 0.3,\\
27.2 $\pm$ 0.3\\
$>$10.0 &Nolan et al.~(2006)& Radar\\

$>15$ & Dykhuis et al.~(2012)& Dust morphology\\

\enddata

\end{deluxetable}

\clearpage
\begin{deluxetable}{lccccl}
%\tabletypesize{\scriptsize}
%\rotate
\tablecaption{Image Information
\label{filters}}
\tablewidth{0pt}
\tablehead{  Filter & $\lambda$\tablenotemark{a}  & FWHM\tablenotemark{b}   &N\tablenotemark{c} & t\tablenotemark{d}  }

\startdata

F475 & 4760 & 1458 &  12&10-20 \\
F555 & 5346 & 1193 & 12 & 20\\
F606 & 5907 & 2342 &  36 & 180\\
F625 & 6318 & 1442 & 24 & 20\\

\enddata

\tablenotetext{a}{The central wavelength of the filter in \AA\ }
\tablenotetext{b}{The full-width-half-maximum of the filter in \AA\ according to the Space Telescope Science Institutes instrument handbook }
\tablenotetext{c}{Number of images}
\tablenotetext{d}{Integration time, seconds}

\end{deluxetable}

\clearpage

\begin{deluxetable}{clcrrrccccr}
%\tabletypesize{\scriptsize}
%\rotate
\tablecaption{Observing Geometry 
\label{geometry}}
\tablewidth{0pt}
\tablehead{  UT Date\tablenotemark{a}   & $\Delta T_p$\tablenotemark{b} & $\nu$\tablenotemark{c} & \colhead{$r_H$\tablenotemark{d}}  & \colhead{$\Delta$\tablenotemark{e}} & \colhead{$\alpha$\tablenotemark{f}}   & \colhead{$\theta_{-\odot}$\tablenotemark{g}} &   \colhead{$\theta_{-V}$\tablenotemark{h}}  & \colhead{$\delta_{\oplus}$\tablenotemark{i}}   }
\startdata

2006 Apr 10 &-60& 2.9 &1.240& 0.292 & 31.2 & 222.1 & 290.6& -30.8\\
2006 Apr 11&-59& 2.9 &1.231& 0.283 & 31.8 & 221.0 &290.4& -31.5\\

\enddata

%% Text for table notes should follow after the \enddata but before
%% the \end{deluxetable}. Make sure there is at least one \tablenotemark
%% in the table for each \tablenotetext.

\tablenotetext{a}{Date of observation at 22:00 UTC}
\tablenotetext{b}{Number of days from perihelion (UT 2006-Jun-09) Negative numbers indicate pre-perihelion observations}
\tablenotetext{c}{True anomaly, in degrees}
\tablenotetext{d}{Heliocentric distance, in AU}
\tablenotetext{e}{Geocentric distance, in AU}
\tablenotetext{f}{Phase angle, in degrees}
\tablenotetext{g}{Position angle of the projected anti-Solar direction, in degrees}
\tablenotetext{h}{Position angle of the projected negative heliocentric velocity vector, in degrees}
\tablenotetext{i}{Angle between Earth and target orbital plane, in degrees}

\end{deluxetable}

\clearpage

\begin{deluxetable}{lccccl}
%\tabletypesize{\scriptsize}
%\rotate
\tablecaption{Nucleus and Coma Colors\tablenotemark{a}
\label{colors}}
\tablewidth{0pt}
\tablehead{  Color & Nucleus (0 - 27 km)  & Coma (27 km - 136 km)& Coma (136 - 245 km)  }

\startdata

F475-F555 & 0.43$\pm$0.04 & 0.42$\pm$0.18 &0.42$\pm$0.26 \\
F555-F606 & 0.25$\pm$0.04 & 0.26$\pm$0.05  & 0.26$\pm$0.06\\
F555-F625 & 0.47$\pm$0.04 & 0.49$\pm$0.04 & 0.49$\pm$0.06\\

\enddata

\tablenotetext{a}{In magnitudes}

\end{deluxetable}

\clearpage
\begin{deluxetable}{llcccccccc}
%\tabletypesize{\scriptsize}
%\rotate
\tablecaption{73P-C Nucleus and Coma Lightcurve Phase and Phase Shift
\label{params34}}
\tablewidth{0pt}
\tablehead{  Aperture& $\theta_i$\tablenotemark{a} & $\theta_o$\tablenotemark{b} & $\theta_e$\tablenotemark{c} & $\mid \Delta$ Phase $\mid$\tablenotemark{d} }

\startdata

1& 0 & 27  & 13 & 0\\
2&27 & 136  & 82 &0.21 $\pm$ 0.05\\
3&136& 245 & 190 & 0.46 $\pm$ 0.08\\
\enddata

\tablenotetext{a}{Inner radius of annulus in km, 1$\arcsec \sim $ 209 km}
\tablenotetext{b}{Outer radius of annulus in km}
\tablenotetext{c}{Effective radius of annulus in km}
\tablenotetext{d}{Phase lag relative to the lightcurve of the nucleus (hours).}
\end{deluxetable}

\clearpage

\clearpage

\clearpage

\clearpage

\clearpage

\clearpage

\clearpage

\end{document}